\documentclass[conference]{IEEEtran}
\IEEEoverridecommandlockouts
\usepackage{cite}
\usepackage{amsmath,amssymb,amsfonts}
\usepackage{algorithmic}
\usepackage{graphicx}
\usepackage{textcomp}
\usepackage{makecell}
\usepackage{xcolor}
\usepackage{hyperref}
\usepackage{adjustbox}
\usepackage{balance}
\def\BibTeX{{\rm B\kern-.05em{\sc i\kern-.025em b}\kern-.08em
    T\kern-.1667em\lower.7ex\hbox{E}\kern-.125emX}}
\begin{document}

\title{Hacking VMAF and VMAF NEG: vulnerability to different preprocessing methods}

% \author{\IEEEauthorblockN{1\textsuperscript{st} Maksim Siniukov}
% \IEEEauthorblockA{\textit{Graphics and Media Lab} \\
% \textit{Lomonosov Moscow State University}\\
% Moscow, Russia \\
% maksim.siniukov@graphics.cs.msu.ru}
% \and
% \IEEEauthorblockN{2\textsuperscript{nd} Anastasia Antsiferova}
% \IEEEauthorblockA{\textit{Graphics and Media Lab} \\
% \textit{Lomonosov Moscow State University}\\
% Moscow, Russia \\
% aantsiferova@graphics.cs.msu.ru}
% \and
% \IEEEauthorblockN{3\textsuperscript{rd} Dmitriy Kulikov}
% \IEEEauthorblockA{\textit{Graphics and Media Lab} \\
% \textit{Lomonosov Moscow State University, Dubna State University}\\
% Moscow, Russia \\
% dkulikov@graphics.cs.msu.ru}
% \and
% \IEEEauthorblockN{4\textsuperscript{th} Dmitriy Vatolin}
% \IEEEauthorblockA{\textit{Graphics and Media Lab} \\
% \textit{Lomonosov Moscow State University}\\
% Moscow, Russia \\
% dmitriy@graphics.cs.msu.ru}
% }
\maketitle

\begin{abstract}
Video-quality measurement plays a critical role in the development of video-processing applications. In this paper, we show how video preprocessing can artificially increase the popular quality metric VMAF and its tuning-resistant version, VMAF NEG. We propose a pipeline that tunes processing-algorithm parameters to increase VMAF by up to 218.8\%. A subjective comparison revealed that for most preprocessing methods, a video’s visual quality drops or stays unchanged. We also show that some preprocessing methods can increase VMAF NEG scores by up to 23.6\%.
\end{abstract}


\section{Introduction}
Video-quality measurement is critical to the development of video-processing algorithms such as compression, video enhancement, adaptive network streaming, and super-resolution. Despite the existence of objective quality metrics, visual quality is often a primary means of testing new algorithms. Perceptual quality is measurable through subjective evaluation, yielding the most-accurate results, but because subjective evaluation is complex and time consuming, it is unsuitable for frequent use; objective quality metrics are more applicable in practice. Yet to ensure precision, not just speed, metrics must correlate highly with perceptual results. Today, a popular and widespread full-reference metric is the Video Multi-method Fusion Approach (VMAF), developed by Netflix in cooperation with the University of Southern California. VMAF is machine-learning based and combines multiple elementary video features (VIF\cite{b1}, DLM\cite{b2} at different scales, and various TI measures) with the machine-learning-based Support Vector Regression technique. The regressor is trained using mean-opinion-score (MOS) data to approximate its per-frame values. VMAF has become popular thanks to its high correlation with subjective quality, as proven by independent researchers\cite{b3}. Since the first release, its developers have published two sets of versions: original models (0.6.1, 0.6.2, 0.6.3, and their variations) and a special ``no-enhancement gain'' (NEG) model for comparing video encoders. A NEG model emerged because video preprocessing can artificially increase VMAF; this characteristic makes the original metric vulnerable, and algorithm developers can exploit it to achieve higher values. Our work’s aim is to show that VMAF can be increased both with or without a change in visual quality. We also demonstrate that VMAF NEG is susceptible to preprocessing and can be artificially increased as well. We propose transformations that considerably increase VMAF and VMAF NEG, in addition to providing results from a subjective evaluation in which the perceptual quality of VMAF-tuned preprocessed videos either declines or stays unchanged.

\section{Related work}
Tuning for objective metrics is used throughout the industry to achieve better scores in video comparisons. It may, however, decrease the subjective quality, just as tuning for subjective metrics may decrease objective quality. The authors of\cite{b4} analyzed the effect of video preprocessing on visual quality and compared it with objective metrics. In their experiments, subjective quality improved when the Gaussian-blurring PSNR decreased. The results showed that a no-reference metric, NIQE,  correlated better with subjective scores for preprocessed videos. But NIQE can also be inconsistent. In other research\cite{b43}, the authors demonstrated that NIQE has issues when estimating the quality of dark and highly textured frames. Many video-processing algorithms implement tuning filters, which are especially popular for video encoders. For example, x264 and x265 provide tuning options for PSNR and SSIM, and recently, the developers of libaom implemented VMAF tuning \cite{b5}. In their paper, they showed that preprocessing enabled a substantial compression gain (the bitrate fell by 37.91\% for the same VMAF), while rate-distortion optimization enabled a much smaller gain (4.69\%). The authors of ``Hacking VMAF with Video Color and Contrast Distortion'' \cite{b6} observed that histogram equalization and unsharp masking can artificially increase VMAF by 5–6\%, and other researchers also proved VMAF’s vulnerability to preprocessing\cite{b6_5_Ozer}. To separate preprocessing enhancement from pure compression loss, Netflix introduced a new metric model called VMAF No Enhancement Gain (NEG)\cite{b7}. It recommends using the NEG mode to evaluate codecs and the ``default'' mode to concurrently assess compression and enhancement. No independent researchers have yet published an analysis of VMAF NEG.

\section{Proposed method}
\begin{figure}[htbp]
\centerline{\includegraphics[width=\linewidth]{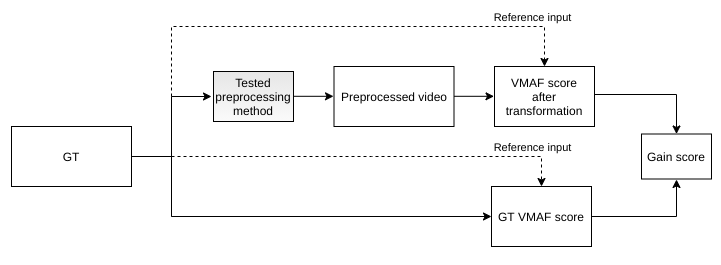}}
\caption{Proposed tuning pipeline without compression.}
\label{fig_pipline1}
\end{figure}

\begin{figure}[htbp]
\centerline{\includegraphics[width=\linewidth]{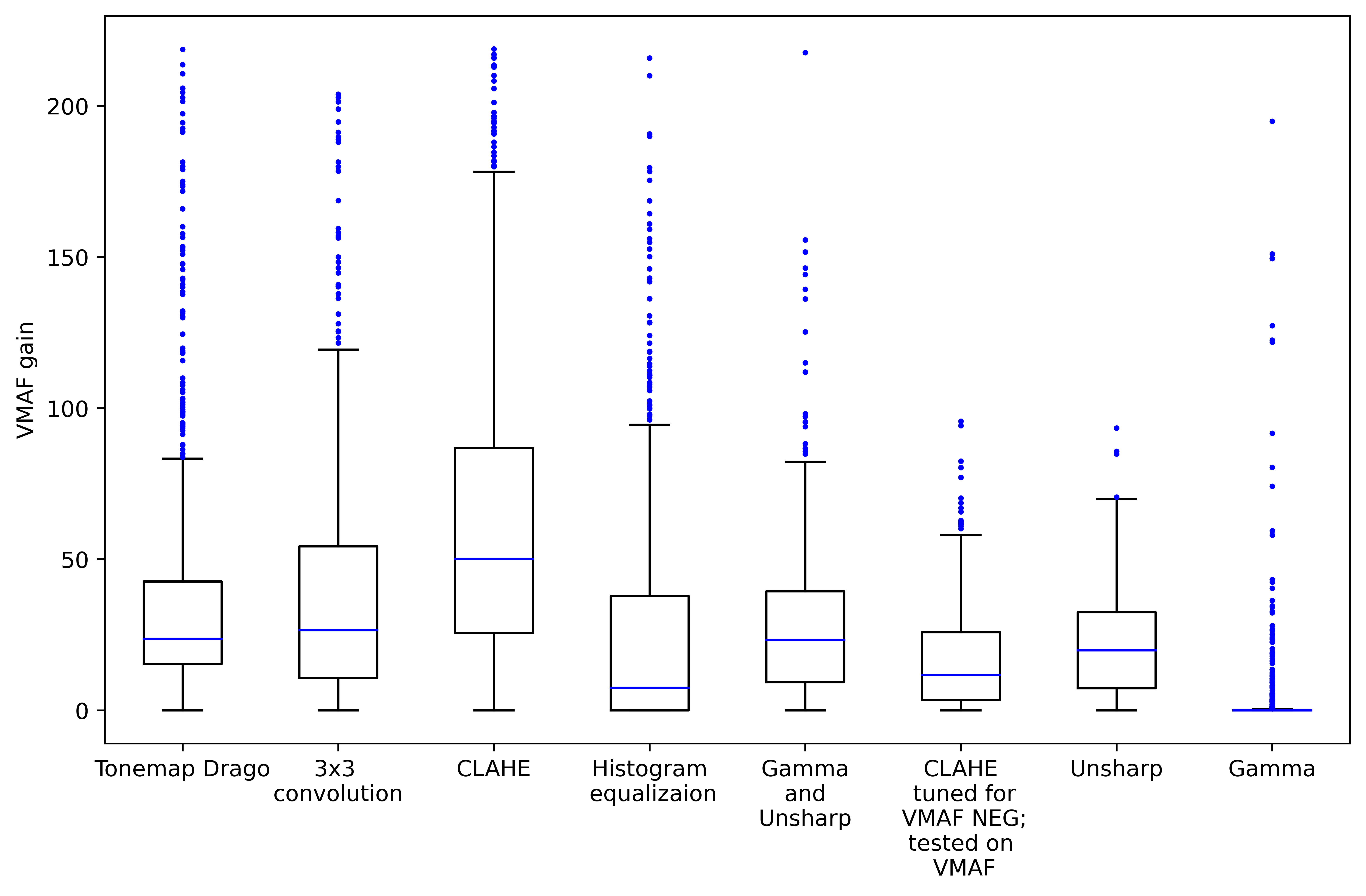}}
\caption{VMAF gain for different preprocessing methods. For examples with negative gain no preprocessing was applied, so they have a zero gain for statistics.}
\label{figbox}
\end{figure}

\begin{figure*}[htbp]
\centerline{\includegraphics[width=\linewidth]{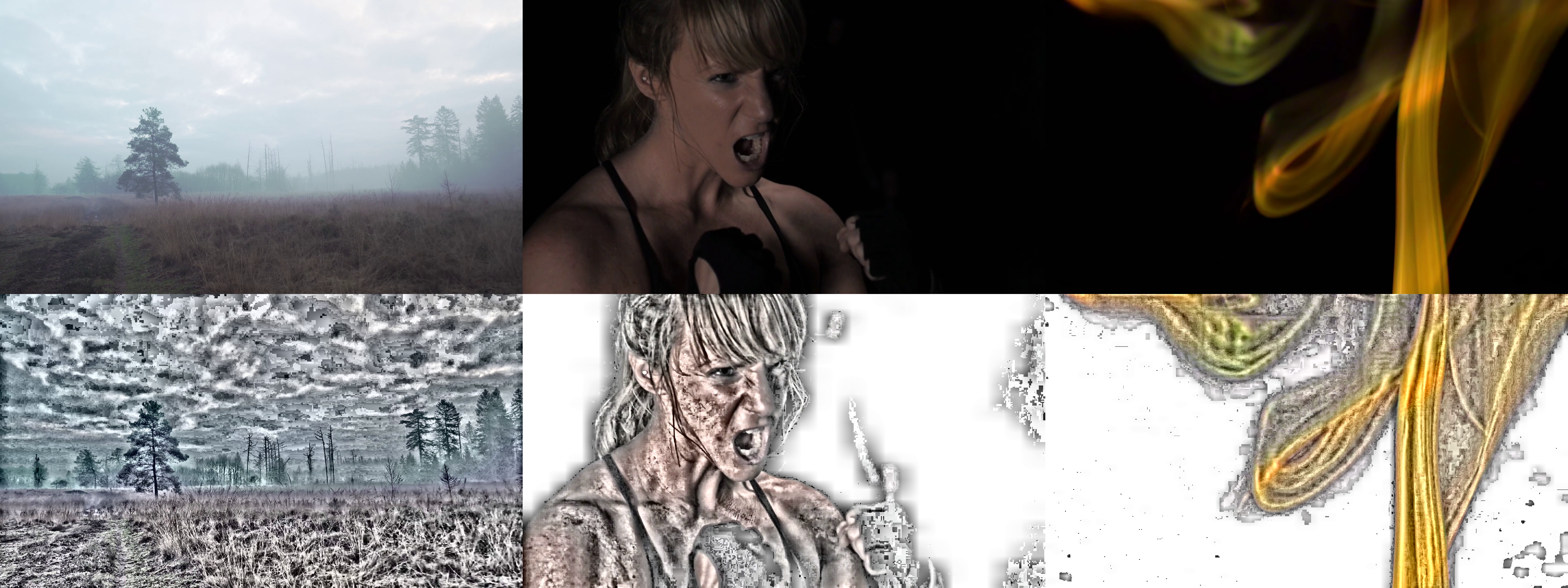}}
\caption{Comparison of frames from original videos (top) and videos preprocessed using the CLAHE method (bottom). VMAF increased by 181.22\% for the video at left, by 147.30\% for the one in the middle video, and by 114.75\% for the one at right.}
\label{fig3ex}
\end{figure*}

Fig.~\ref{fig_pipline1} depicts our pipeline for studying how various transforms affect VMAF. We transformed the input video using various preprocessing methods, then calculated the VMAF score between the preprocessed and original (``ground truth'', or GT) video sequences. Next we compared the resulting VMAF score with the VMAF measured between two GT videos (as it can differ from 100 for two instances of the same video, the value is usually around 100 \cite{b8}). If the resulting VMAF score was lower than the score for videos without preprocessing, an original video was kept and zero gain was accounted in statistics. 
We tested several preprocessing methods: tone-mapping algorithms (Drago’s tone-mapping method \cite{b9}, Mantiuk’s tone-mapping method \cite{b10}, Retinex, and Reinhard’s tone-mapping method \cite{b11}), contrast-limited adaptive histogram equalization (CLAHE), histogram equalization, contrast-based transformations (gamma correction, three-degree polynomial contrast transformation, and linear contrast transformation), and unsharp masking. Our testing included a combination of gamma correction and unsharp masking, as well as a CLAHE tuned for VMAF NEG to increase the VMAF value. In addition to the methods above, we applied stochastic gradient descent to approximate convolution weights. (We slightly increased the value of each cell, calculating the resulting VMAF value each time. Then we found  gradient-vector-approximation decomposition for VMAF on each kernel weight using an approximate right-derivative formula.) To ensure the approaches we considered were suitable for real-world data, we constructed a data set containing professional camera footage, user-generated sequences, and computer-generated imagery. The total number of videos was 618. Since VMAF models are designed for FullHD resolution, all videos were FullHD. We calculated the VMAF score for the entire video as the mean of all per-frame VMAF values. Accordingly, varying the video lengths did not affect VMAF judgment, so we used only the first 10 frames to find the best preprocessing method. Most of the methods we tested had several parameters, and each parameter raised exponentially the computational complexity of an optimal-combination search. To reduce the number of tested parameters and accelerate convergence, our approach used the $(\mu + \lambda)$ genetic algorithm \cite{b12}. Fig.~\ref{figbox} shows the methods that yielded the highest gain values (e.g., the best tone-mapping method and best trained convolution).\\
 The results showed that CLAHE yielded the best VMAF increase. The third quartile (top and bottom lines of the boxes in Fig.~\ref{figbox}) was an 86.8\% gain, and the median was 50.2\%; the highest gain was 218.8\%. In most cases where the VMAF score increased considerably (more than 50\%), the subjective quality decreased. Examples of this trend appear in Fig.~\ref{fig3ex}. The VMAF score increased even when applying simple histogram equalization without tunable parameters—in particular, the third-quartile gain was 37.9\%.

\section{Results verification on encoded streams}
\begin{figure}[htbp]
\centerline{\includegraphics[width=\linewidth]{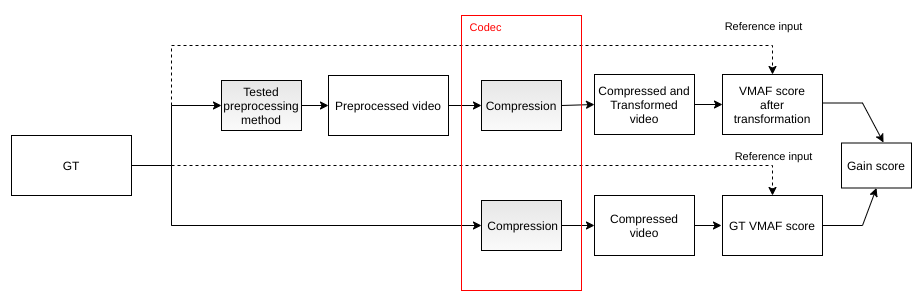}}
\caption{Tuning pipeline with compression.}
\label{figpipline2}
\end{figure}

\begin{figure}[htbp]
\centerline{\includegraphics[width=\linewidth]{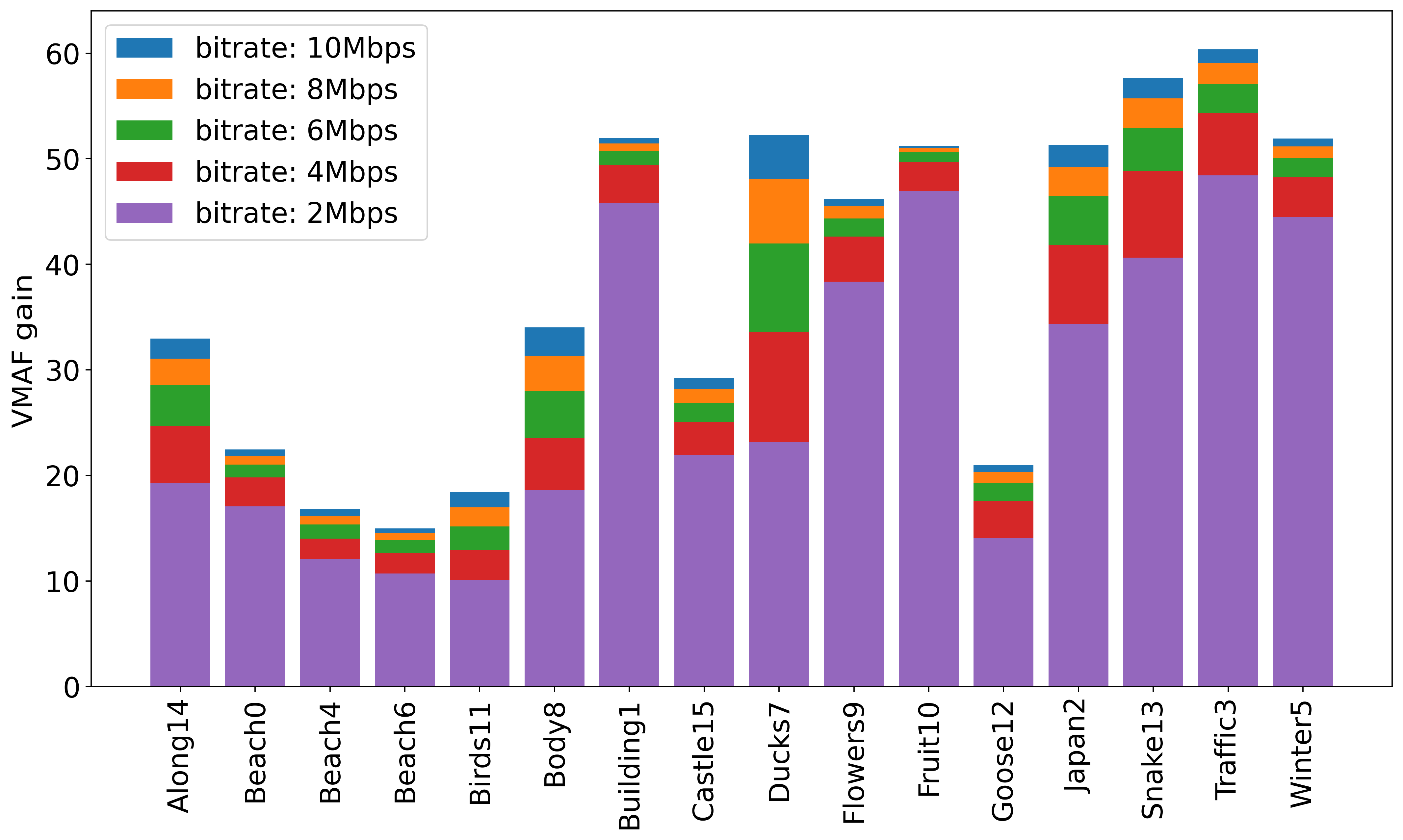}}
\caption{VMAF gain at different bitrates. We tuned the first 10 frames of each video, then used x265 to compress the full 300 frames. Gains are generally higher at high bitrates.}
\label{figcodec}
\end{figure}

\begin{figure*}
\centering{\includegraphics[trim={0 0 0 3cm},clip,width=\linewidth]{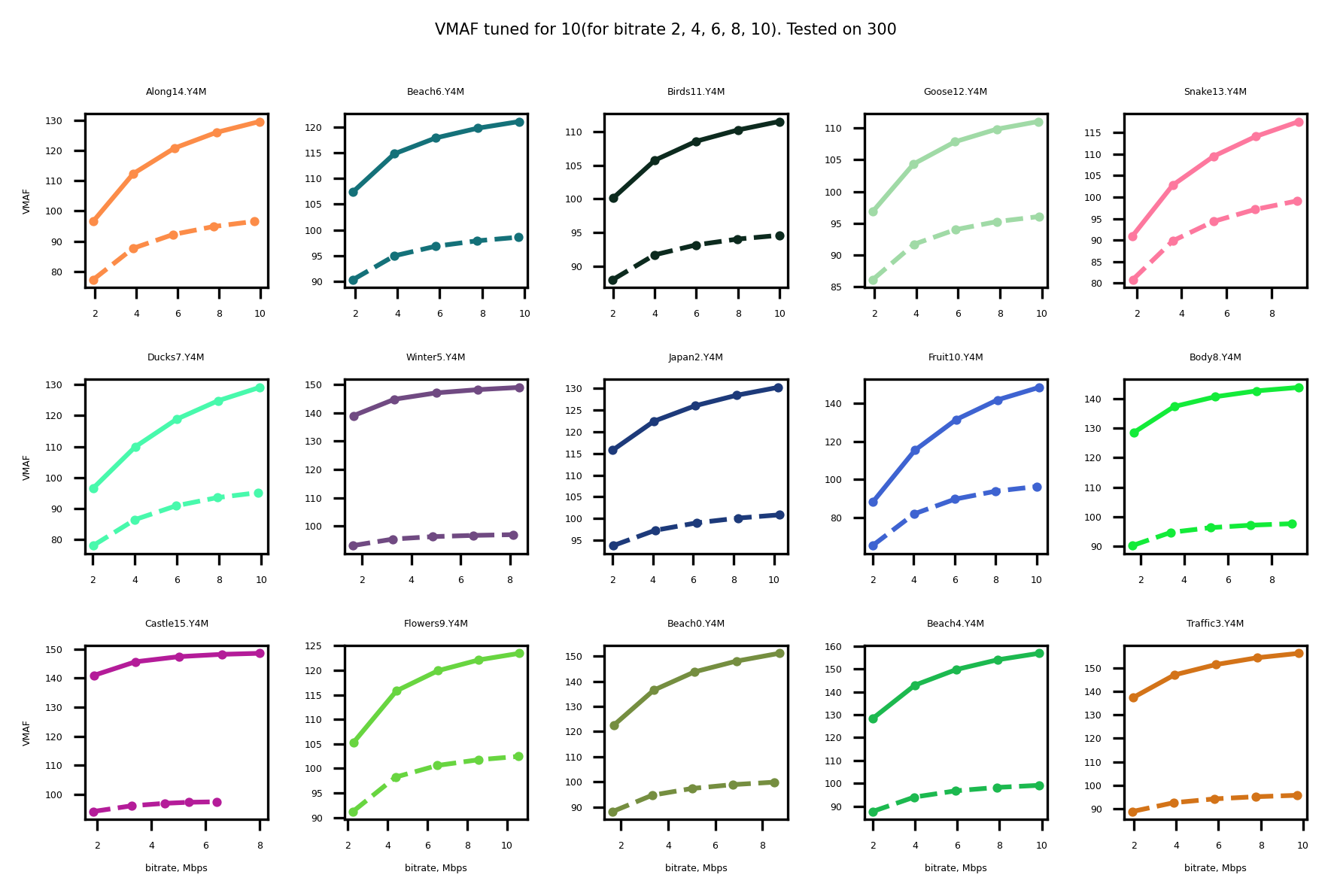}}
\caption{Rate-distortion curves for original- and preprocessed-video compression. Each dashed line is for the GT video, and the solid line is for the preprocessed video. The VMAF increase due to preprocessing is generally higher at high bitrates.}
\label{fig_RD}
\end{figure*}

VMAF is widely used in codec comparisons as a main objective estimate of video quality—for example, in\cite{b13}. To ensure our proposed approach is applicable in this field, we changed the main comparison pipeline by adding a video-compression step before calculating VMAF, as Fig. ~\ref{figpipline2} shows. 
Our approach therefore calculated VMAF twice: once for the compressed GT video, then once for the compressed and preprocessed video. The plot below shows the difference between corresponding VMAF scores (compressed and compressed with preprocessing). We used x265 to compress the videos at the following bitrates: 2, 4, 6, 8, and 10 Mbps. We increased all video durations to 300 frames to allow codec stabilization. The results for several videos from the data set appear in Fig. ~\ref{figcodec} and Fig. ~\ref{fig_RD}. For each one, the VMAF gain stayed the same relative to the tuned but uncompressed video. The VMAF generally increased more at high bitrates.

\subsection{Subjective evaluation}
To analyze how video preprocessing (which increases VMAF scores) affects perceptual quality, we conducted a subjective comparison. We chose six representative videos that differ in spatial and temporal complexity as well as in brightness. The crowdsourced Subjectify.us service allowed us to receive an assessment from 114 participants. These participants were to look at pairs of images processed using different methods and then choose from each pair the image with the best subjective quality. Votes from those who failed to correctly answer the verification questions (which exhibited obvious quality differences) were excluded. The total number of responses was 1,584.\\

\begin{figure}[htbp]
\centerline{\includegraphics[trim={0 0 3cm 0},clip,width=\linewidth]{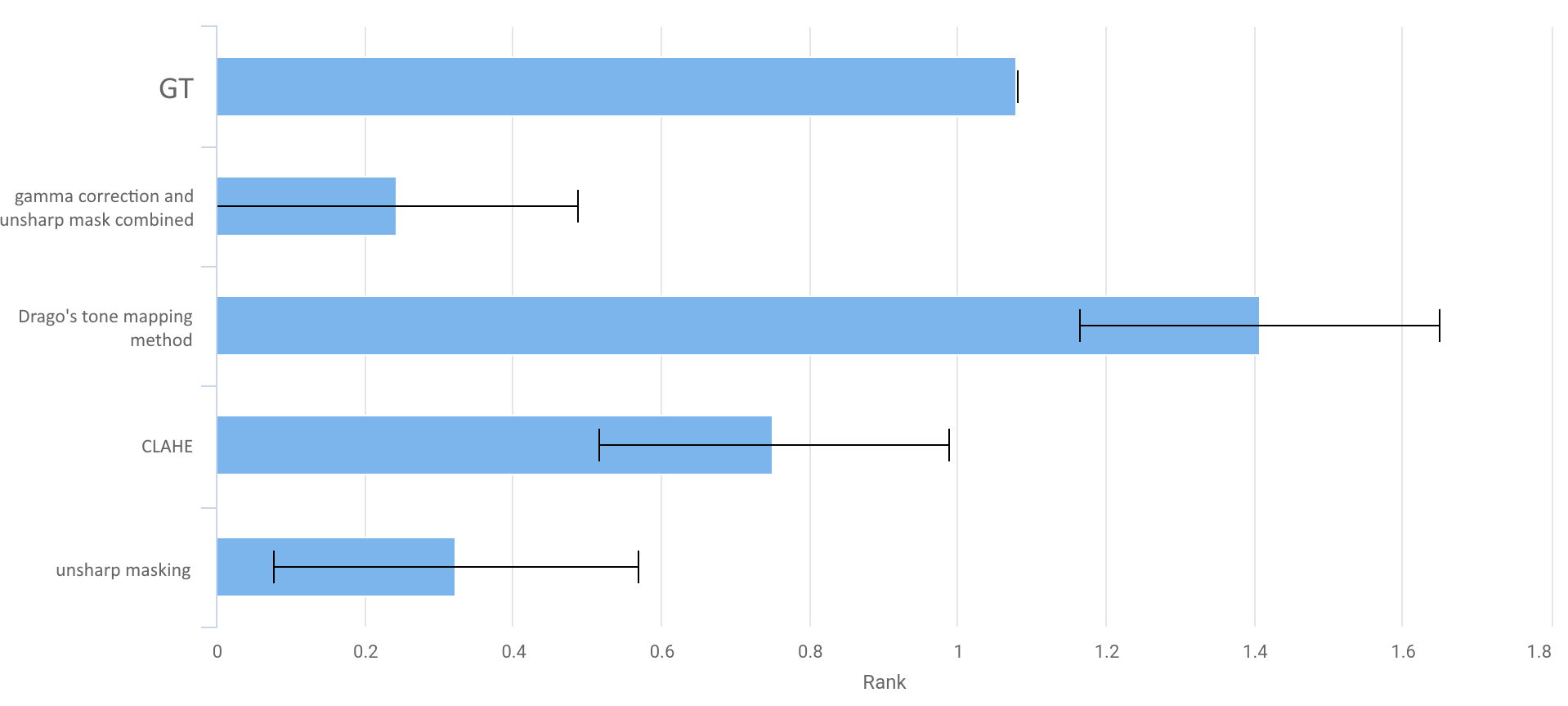}}
\caption{Subjective-evaluation results. The ranks came from a Bradley-Terry model for pairwise comparisons.}
\label{figSubj}
\end{figure}

The comparison included six versions of preprocessed videos: GT (original), gamma correction and unsharp masking combined, Drago’s tone-mapping method, CLAHE, gamma, and unsharp masking. In a direct comparison, GT proved to be better than the preprocessed video in 58.3\% of cases, meaning they were for the most part visually indistinguishable. In all the videos, however, VMAF rose artificially by a mean of 46\%. The combination of gamma correction and unsharp masking provided the worst visual quality. The highest subjective-evaluation values went to Drago’s tone-mapping algorithm. Fig. ~\ref{figSubj} shows the results of the comparison.

Table~\ref{tab1} provides the values of VMAF gain by tested preprocessing methods. Drago's tone-mapping method provided similar gain on average compared to unsharp masking and gamma+unsharp masking and much better visual quality.

\begin{table}[htbp]
\caption{VMAF increases by tuned preprocessing method}
\begin{center}
\begin{adjustbox}{width=\columnwidth,center}
\begin{tabular}{|c|c|c|c|c|}
\hline
&\multicolumn{4}{|c|}{Preprocessing method} \\
\cline{2-5} 
\makecell{Video\\ name} & \makecell{ Gamma correction\\ and \\ Unsharp masking\\ combination }& \makecell{Drago’s tone\\ mapping method}& CLAHE & \makecell{Unsharp\\ masking} \\
\hline
Along & 11.39 & 19.69 & 41.62 & 7.92  \\
\hline
Beach & 18.41 & 16.02 & 34.27 & 18.4 \\
\hline
Body & 5.6 & 20.84 & 45.83 & 5.6 \\
\hline
Building & 64.9 & 28.13 & 52.83 & 65.03 \\
\hline
Castle & 22.17 & 24.21 & 42.63 & 22.22 \\
\hline
Ocean & 85.22 & 107.09 & 191.3 & 85.22 \\
\hline
Avg. & 34.62 & 37.00 & 68.08 &34.07 \\
\hline
\end{tabular}
\end{adjustbox}
\label{tab1}
\end{center}
\end{table}

\subsection{VMAF NEG analysis}
\begin{figure}[htbp]
\centerline{\includegraphics[width=\linewidth]{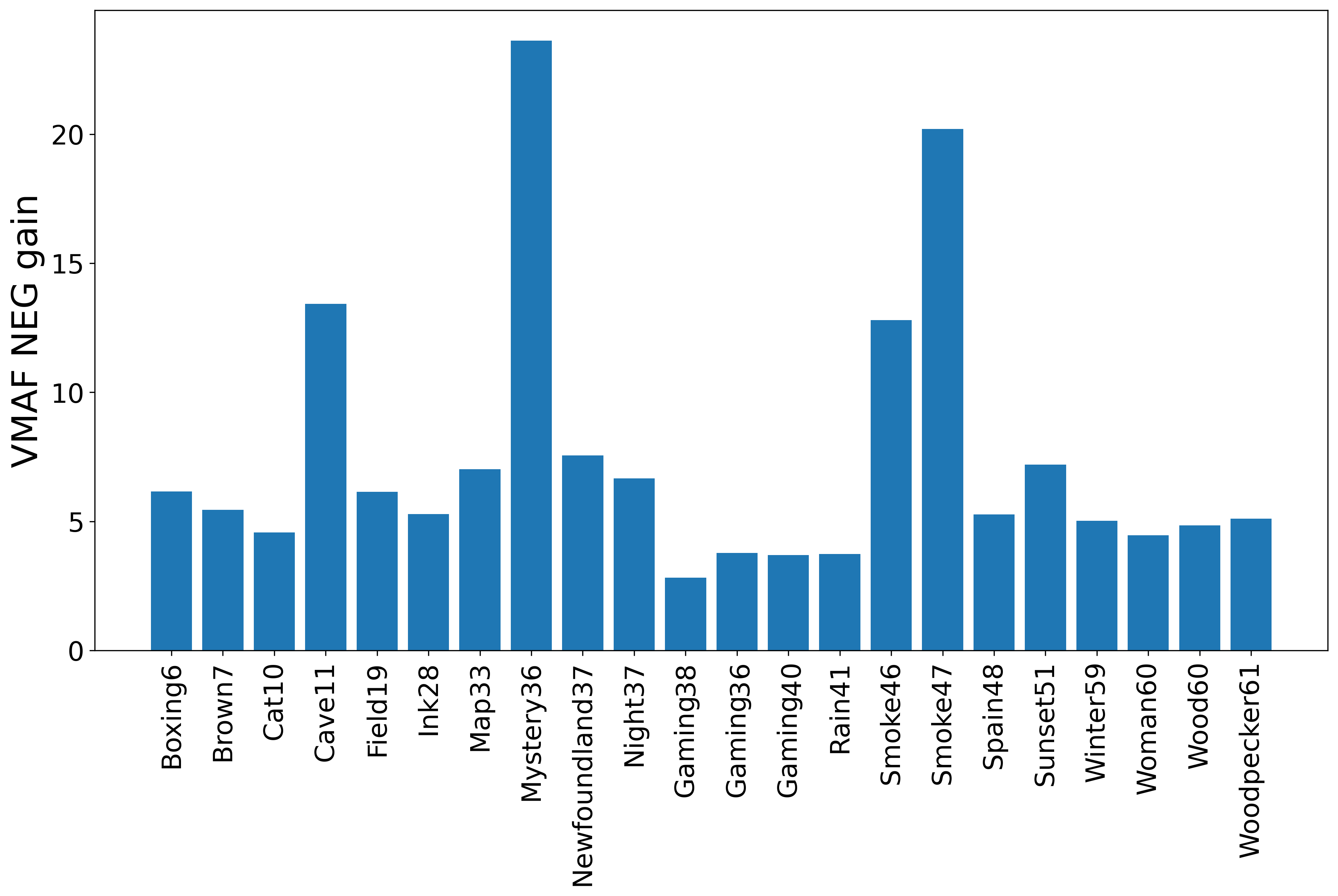}}
\caption{VMAF NEG gain for different videos. The preprocessing method was gamma correction plus unsharp masking.}
\label{fig_neg}
\end{figure}

\begin{figure}[htbp]
\centerline{\includegraphics[width=\linewidth]{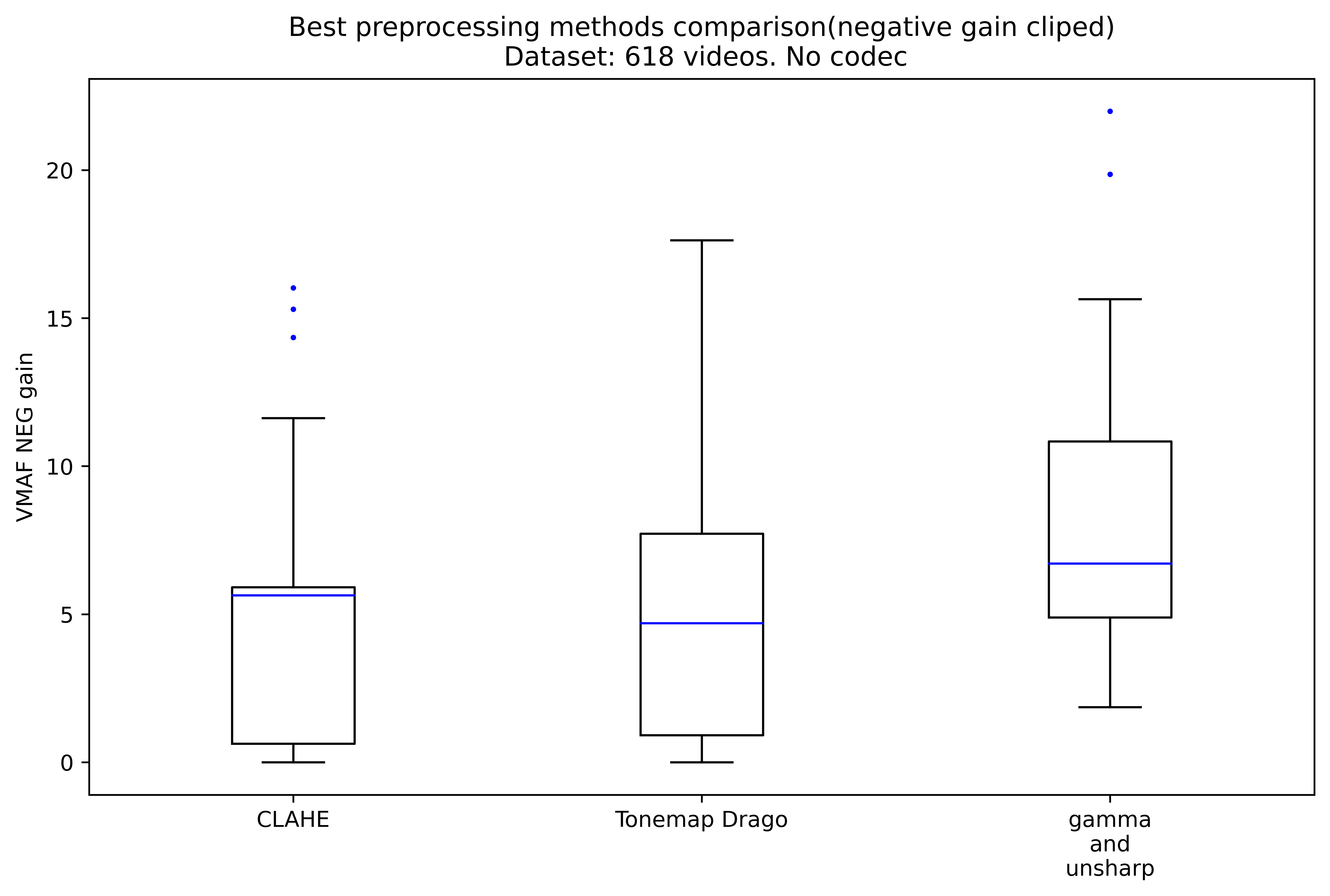}}
\caption{VMAF NEG gain for different preprocessing methods. For examples with negative gain no preprocessing was applied, so they have a zero gain for statistics.}
\label{fig_neg_boxplot}
\end{figure}

\begin{figure*}[!htbp]
\centerline{\includegraphics[width=\linewidth]{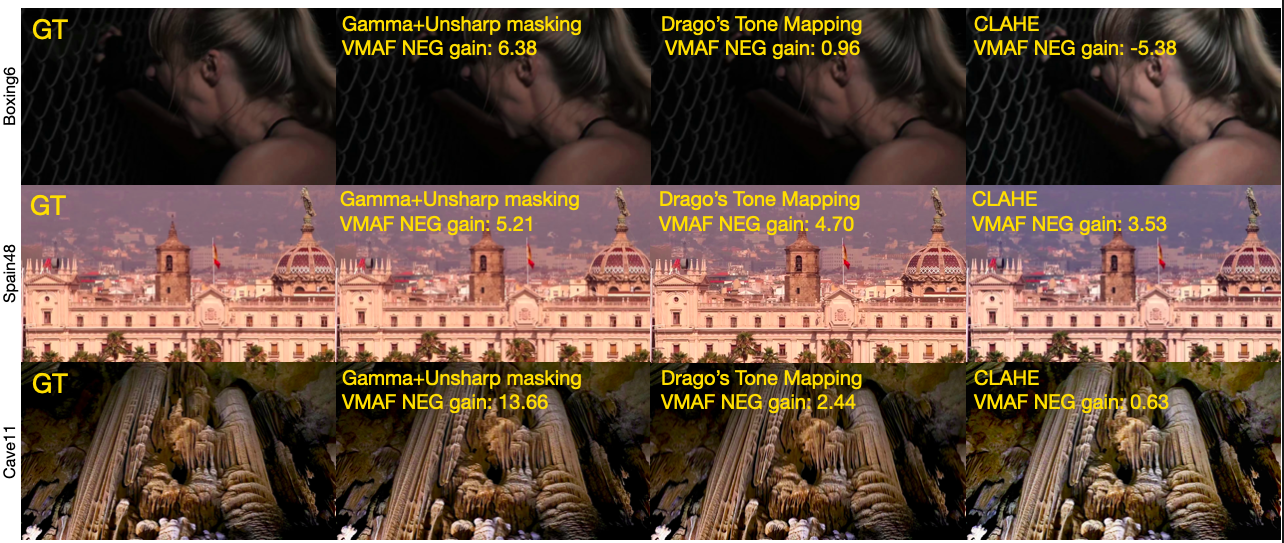}}
\caption{Examples of frames with different preprocessing methods, tunned for VMAF NEG increase. Metric gain is calculated as a difference between VMAF NEG score for preprocessed and compressed video, and only compressed video.}
\label{images_neg}
\end{figure*}

To explore the impact of the above-described transformations on VMAF NEG, we used the pipeline from Fig. ~\ref{figpipline2} and replaced VMAF with VMAF NEG. The combination of gamma correction and unsharp masking yielded the greatest increase in VMAF NEG: 23.6\%. The gain exceeded 5\% in 15 out of 22 videos (Fig. ~\ref{fig_neg}), meaning video preprocessing still affects the NEG version. Fig. ~\ref{fig_neg_boxplot} shows the other methods that yielded the high gain values. However, visual quality of VMAF NEG-tunned preprocessed videos is better, that of VMAF-tunned preprocessing. Fig.~\ref{images_neg} shows the examples of frames with and without preprocessing.

\section{Conclusion}
In this paper we described our investigation of methods for increasing VMAF and VMAF NEG. Among the preprocessing methods we tested, CLAHE showed the highest gain for VMAF: 63.17\%. Unsharp masking combined with gamma correction increased VMAF NEG by 5.93\% on average. Subjective evaluation showed that preprocessed videos with greater VMAF mostly have the same or worse visual quality. The best visual quality came by applying tone mapping to the original video.\\

Absent subjective evaluation, VMAF and VMAF NEG should not be the only video-quality measure in a benchmark, because basic video-processing methods can increase them. We demonstrated that these methods can boost the VMAF value considerably despite reducing or leaving unchanged the visual quality. Developers of video-processing and compression algorithms can implement in a codec certain tuning methods that may either have no effect or cause visual-quality loss. In addition, video preprocessing can artificially increase VMAF NEG, a valuable metric when comparing codecs, by up to 23.6\%.\\

Measuring subjective quality in video comparisons may also provide controversial results, however. In our tests, tone mapping applied to the original video ranked higher in a visual comparison. Because viewers tend to choose the most visually appealing image, objective full-reference metrics are also necessary to check picture similarity with reference. Thus, fairer and more-reliable comparisons require a combination of different analysis methods and professional benchmarks.

\balance
\section*{Acknowledgment}

This work was partially supported by the Russian Foundation for Basic Research under Grant 19-01-00785a and by the ``Intellect'' noncommercial fund for science and educational development.

\end{document}